\documentstyle[aps,preprint,floats,tighten,epsf]{revtex}

\begin{document}
\preprint{\vbox{\hbox{JLAB-THY-98-05}
                \hbox{TRI--PP--98--03}
                \hbox{hep-lat/9803004}}}

\title{\boldmath$O(1/M^3)$ effects for heavy-light mesons in lattice NRQCD}

\author{Randy Lewis}
\address{Jefferson Lab, 12000 Jefferson Avenue, Newport News, VA, U.S.A. 23606 
         \\ {\rm and} \\
       Department of Physics, University of Regina, Regina, SK, Canada S4S 0A2}
\author{R. M. Woloshyn}
\address{TRIUMF, 4004 Wesbrook Mall, Vancouver, BC, Canada V6T 2A3}

\date{February 1998}
\maketitle

\begin{abstract}
The masses of $^1S_0$ and $^3S_1$ mesons containing a single heavy quark 
are computed in the quenched approximation.
The light quark action and gauge field action are both classically-improved and 
tadpole-improved, and the couplings to the heavy quark are organized by the
$1/M$ expansion of tadpole-improved NRQCD.
At each of two lattice spacings, near 0.22fm and 0.26fm,  meson masses are
obtained for heavy quarks spanning the region between charmed and bottom 
mesons.
Results up to $O(1/M)$, $O(1/M^2)$ and $O(1/M^3)$ are displayed separately,
so that the convergence of the heavy quark expansion can be discussed.
Also, the effect of each term in the $O(1/M^3)$ contribution is computed
individually.  For bottom mesons the $1/M$-expansion appears to be 
satisfactory, but the situation for charmed mesons is less clear.
\end{abstract}


\section{INTRODUCTION}

Long-distance, nonperturbative QCD interactions can be studied numerically
by discretizing space-time, if the lattice spacing ``$a$'' is sufficiently
small to allow a matching to perturbative QCD.  To include the effects of
a quark whose inverse mass is smaller than the lattice spacing, it is
natural to use an effective Lagrangian which is ordered by powers of the
inverse quark mass.  Two different notations which have been used for this
expansion are heavy quark effective theory (HQET)\cite{HQET} and 
nonrelativistic QCD (NRQCD)\cite{NRQCD,NRQCD2}.

In the present work, the NRQCD formalism is used to study the masses of
the ground state (${}^1S_0$) and first excited state (${}^3S_1$) heavy-light 
mesons, i.e. mesons containing a single heavy quark.  
These masses are well-known experimentally\cite{PDG},
and have been previously determined from lattice 
NRQCD\cite{review,Japan,previous,ADCSS,melbourne} up to
$O(1/M^2)$.  The primary goal of the present research is to extend the
calculation to $O(1/M^3)$, and to display the effects of each new term
individually.  This provides an indication of the convergence of the 
$1/M$ expansion of lattice NRQCD.  Of particular interest are the 
physically-relevant cases of charmed and bottom mesons, both of which will
be discussed herein.

To allow the use of coarsely-spaced lattices, the light-quark and gauge
terms in the action will be classically-improved, and the entire action will
be tadpole-improved.  Not only does a larger lattice spacing imply speedier
simulations, but it also means that the minimum heavy quark mass of NRQCD 
(which is of the order of the inverse lattice spacing) is reduced.
This opens the possibility of using NRQCD to study charmed mesons.

Lattice NRQCD has been used extensively for studies of quarkonium\cite{QQbar},
where the $1/M$ expansion is replaced by a velocity expansion,
and it was concluded that the velocity expansion for the spin splitting of 
charmonium S-waves does not converge very quickly.\cite{Trottier}  
In the present work, the spin splittings of S-wave heavy-light charmed mesons
nicely satisfy $|O(1/M^3)| < |O(1/M^2)| < |O(1/M)|$.  However, there is an
individual $O(1/M^3)$ term which is larger in magnitude than the total
$O(1/M^2)$ contribution.  Future studies
may be able to improve this situation, and various suggestions
present themselves as conclusions to this exploratory study.

\section{ACTION}

The lattice action can be written as the sum of three terms,
\begin{equation}
   S = S_G(U) + S_q(\bar{q},q;U) + S_Q(\bar{Q},Q;U)~,
\end{equation}
where $U$, $q$ and $Q$ are the gauge field, light quark field and
heavy quark field, respectively.

Following the work of L\"uscher and Weisz\cite{LusWei85},
a gauge field action which is classically-correct up to $O(a^4)$ errors can be 
written by including a sum over $1 \times 2$ rectangular plaquettes ($U_{rt}$) 
as well as $1 \times 1$ elementary plaquettes ($U_{pl}$),
\begin{equation}
   S_G(U) = \frac{\beta}{3}{\rm ReTr}\left[\sum_{pl}(1-U_{pl})
        -\frac{1}{20U_0^2}\sum_{rt}(1-U_{rt})\right]~.
\end{equation}
A tadpole factor, defined by
\begin{equation}
   U_0 = \langle\frac13\mbox{Re}\mbox{Tr}U_{pl}\rangle^{1/4}
\end{equation}
has been introduced to absorb the lattice tadpole effects and thereby improve
the matching to perturbation theory\cite{LepMac}.

A light quark action, with classical errors at $O(a^2)$ in spectral quantities,
has been constructed by Sheikholeslami and Wohlert\cite{SW},
\begin{eqnarray}
   S_q(\bar{q},q;U) &=& -\sum_{x}\bar{q}(x)q(x) \nonumber \\
        && +\kappa\sum_{x,\mu}\left[
           \bar{q}(x)(1-\gamma_{\mu})U_{\mu}(x)q(x+\mu)
           +\bar{q}(x+\mu)(1+\gamma_{\mu})
           U_{\mu}^{\dagger}(x)q(x)\right] \nonumber \\
        && -\frac{g\kappa}{2U_0^3}\sum_{x,\mu,\nu}\bar{q}(x)\sigma_{\mu\nu}
           F_{\mu\nu}(x)q(x)~.
\end{eqnarray}
Again, the tadpole factor has been included for the
reduction of quantum discretization errors.
The lattice field strength tensor is given by
\begin{eqnarray}
  gF_{\mu\nu}(x)&=&\frac{1}{2i}\left(\Omega_{\mu\nu}(x)-\Omega^\dagger_{\mu\nu}
                   (x)\right) - \frac{1}{3}{\rm Im}\left({\rm Tr}\Omega_{\mu
                   \nu}(x)\right)~, \\
   \Omega_{\mu\nu} &=& \frac{-1}{4}\left[
   U_\mu(x)U_\nu(x+\hat\mu)U_\mu^\dagger(x+\hat\nu)U_\nu^\dagger(x) \right.
       \nonumber \\
       && ~~+U_\nu(x)U_\mu^\dagger(x-\hat\mu+\hat\nu)U_\nu^\dagger(x-\hat\mu)
       U_\mu(x-\hat\mu) \nonumber \\
       && ~~+U_\mu^\dagger(x-\hat\mu)U_\nu^\dagger(x-\hat\mu-\hat\nu)
       U_\mu(x-\hat\mu-\hat\nu)U_\nu(x-\hat\nu) \nonumber \\
       && ~~\left.
       +U_\nu^\dagger(x-\hat\nu)U_\mu(x-\hat\nu)U_\nu(x+\hat\mu-\hat\nu)
       U_\mu^\dagger(x) \right]~.
\end{eqnarray}

For quarkonium, the form of the heavy quark action has been discussed in
detail by Lepage et al.\cite{NRQCD2}.
It is convenient to write the heavy quark action in terms of the Hamiltonian,
$H$,
\begin{equation}
   S_Q(\bar{Q},Q;U) = \int{\rm d}^4x\,Q^\dagger(x)(iD_t-H)Q(x)~.
\end{equation}
To discuss heavy-light mesons, it is appropriate to reorganize the velocity 
expansion of $H$, discussed in Ref. \cite{NRQCD2}, into an expansion in 
powers of the inverse heavy quark bare mass, $M$,
\begin{eqnarray}
H &=& H_0 + \delta{H}^{(1)} + \delta{H}^{(2)} + \delta{H}^{(3)} + O(1/M^4) 
      \label{H} \\
H_0 &=& \frac{-\Delta^{(2)}}{2M} \label{H0} \\
\delta{H}^{(1)} &=& -\frac{c_4}{U_0^4}\frac{g}{2M}\mbox{{\boldmath$\sigma$}}
                    \cdot\tilde{\bf B} + c_5\frac{a^2\Delta^{(4)}}{24M}~, \\
\delta{H}^{(2)} &=& \frac{c_2}{U_0^4}\frac{ig}{8M^2}(\tilde{\bf \Delta}\cdot
                    \tilde{\bf E}-\tilde{\bf E}\cdot\tilde{\bf \Delta})
                    -\frac{c_3}{U_0^4}\frac{g}{8M^2}\mbox{{\boldmath$\sigma$}}
               \cdot(\tilde{\bf \Delta}\times\tilde{\bf E}-\tilde{\bf E}\times 
                 \tilde{\bf \Delta}) - c_6\frac{a(\Delta^{(2)})^2}{16nM^2}~, \\
\delta{H}^{(3)} &=& -c_1\frac{(\Delta^{(2)})^2}{8M^3}
                    -\frac{c_7}{U_0^4}\frac{g}{8M^3}\left\{\tilde\Delta^{(2)},
                        \mbox{{\boldmath$\sigma$}}\cdot\tilde{\bf B}\right\}
                    -\frac{c_9}{U_0^8}\frac{ig^2}{8M^3}
                        \mbox{{\boldmath$\sigma$}}\cdot
                        (\tilde{\bf E}\times\tilde{\bf E}
                        +\tilde{\bf B}\times\tilde{\bf B}) \nonumber \\
                 && -\frac{c_{10}}{U_0^8}\frac{g^2}{8M^3}(\tilde{\bf E}^2
                    +\tilde{\bf B}^2)
                    -c_{11}\frac{a^2(\Delta^{(2)})^3}{192n^2M^3}~.
              \label{dH3}
\end{eqnarray}
The coefficients of the Hamiltonian are chosen so the dimensionless 
parameters, $c_i$, are unity at the classical level.  Terms arising from
quantum effects, i.e. containing powers of $g$ unaccompanied
by $\bf E$ or $\bf B$, have not been shown.
A tilde on any quantity indicates that the leading
discretization errors have been removed.  In particular,
\begin{eqnarray}
   \tilde{E}_i &=& \tilde{F}_{4i}~, \\
   \tilde{B}_i &=& \frac{1}{2}\epsilon_{ijk}\tilde{F}_{jk}
\end{eqnarray}
where\cite{NRQCD2}
\begin{eqnarray}
   \tilde{F}_{\mu\nu}(x) &=& \frac{5}{3}F_{\mu\nu}(x) \nonumber \\
                         &-& \frac{1}{6U_0^2}\left[
   U_\mu(x)F_{\mu\nu}(x+\hat\mu)U_\mu^\dagger(x)
   +U_\mu^\dagger(x-\hat\mu)F_{\mu\nu}(x-\hat\mu)U_\mu(x-\hat\mu)
   -(\mu\leftrightarrow\nu)\right]~.
\end{eqnarray}
The various lattice derivatives are defined as follows.
\begin{eqnarray}
   a\Delta_iG(x) &=& \frac{1}{2U_0}[U_i(x)G(x+a\hat\imath)
                    -U^\dagger_i(x-a\hat\imath)G(x-a\hat\imath)] \\
   a\Delta^{(+)}_iG(x) &=& \frac{U_i(x)}{U_0}G(x+a\hat\imath) - G(x) \\
   a\Delta^{(-)}_iG(x) &=& G(x) - 
              \frac{U^\dagger_i(x-a\hat\imath)}{U_0}G(x-a\hat\imath) \\
   a^2\Delta^{(2)}_iG(x) &=& \frac{U_i(x)}{U_0}G(x+a\hat\imath) - 2G(x)
               +\frac{U^\dagger_i(x-a\hat\imath)}{U_0}G(x-a\hat\imath) \\
   \tilde\Delta_i &=& \Delta_i
                   - {a^2\over 6} \Delta^{(+)}_i\Delta_i\Delta^{(-)}_i \\
   \Delta^{(2)} &=& \sum_i \Delta^{(2)}_i \\
   \tilde \Delta^{(2)} &=& \Delta^{(2)} - {a^2 \over 12} \Delta^{(4)} \\
   \Delta^{(4)} &=& \sum_i \left( \Delta^{(2)}_i \right)^2
\end{eqnarray}

All of the terms in Eqs.~(\ref{H0}-\ref{dH3}) appear in Ref. \cite{NRQCD2}
except for the pieces of the $c_9$ and $c_{10}$ terms quadratic in 
$\bf\tilde B$
(because they are of negligibly high order for quarkonium) and the $c_{11}$
term (to be discussed below).  The fact that the Hamiltonian $H$ is complete 
to $O(1/M^3)$ in the classical continuum limit has been shown by 
Manohar\cite{Manohar}.  

It is conventional to separate $H$ into two pieces, $H_0$ and $\delta{H}$,
such that the evolution of a heavy quark Green's function
takes the form\cite{NRQCD2,Gheavy}
\begin{eqnarray}\label{evol1}
G_1 &=& \left(1-\frac{aH_0}{2n}\right)^n\frac{U_4^\dagger}{U_0}
        \left(1-\frac{aH_0}{2n}\right)^n\,\delta_{\vec{x},0} \\
G_{t+1} &=& \left(1-\frac{aH_0}{2n}\right)^n\frac{U_4^\dagger}{U_0}
            \left(1-\frac{aH_0}{2n}\right)^n(1-a\delta{H})G_t~~,~~t>0 
        \label{evol2}
\end{eqnarray}
where $n$ is a parameter which should be chosen to stabilize the numerics.
Notice that for a free heavy quark field, the only relevant terms in the
Hamiltonian are $H_0$ and the terms containing $c_1$, $c_5$, $c_6$ or $c_{11}$.
Setting each of these $c_i$ to its classical value of unity, and working
consistently to $O(1/M^3)$, gives
\begin{equation}
   G_{t+1} = \exp\left[-a\left(H_0-\frac{(\Delta^{(2)})^2}{8M^3}+O(a^3)
             \right)\right]G_t~~,~~ {\rm for~} t>0
\end{equation}
which displays the absence of discretization errors for the free quark
Hamiltonian up to $O(a^2)$.
The terms containing explicit powers of ``$a$'' ($c_5$, $c_6$ and $c_{11}$) 
were added to the Hamiltonian precisely for this purpose.

\section{CORRELATION FUNCTIONS}

A heavy-light meson is created by the following operator,
\begin{equation}
   \sum_{\vec x}Q^\dagger(\vec x)\Gamma(\vec x)q(\vec x)~,
\end{equation}
where $\Gamma(\vec x)$ is a 4$\times$2 matrix containing the spin structure
\begin{eqnarray}\label{gamma1}
   {}^1S_0 &:& \Gamma(\vec x) = (~0~I~)~, \\
   {}^3S_1 &:& \Gamma(\vec x) = (~0~\sigma_i~)~. \label{gamma2}
\end{eqnarray}

Gauge-invariant smearing was also tried according to the method described in
Ref. \cite{Trottier}, but it provided no significant improvement for the
heavy-light S-waves, which already display clear plateaux
for local sources and sinks.

The general form of the meson correlation function is
\begin{equation}
   G_{\rm meson}(\vec p,t) = \sum_{\vec y}{\rm Tr}\left[\gamma_5(M^{-1})
   ^\dagger(\vec y - \vec x)\gamma_5\Gamma^\dagger_{(sk)}(\vec y)G_t(\vec y -
   \vec x)\Gamma_{(sc)}(\vec x)\right]\exp\left(-i\vec p\cdot(\vec y - 
   \vec x)\right)~.
\end{equation}
Because NRQCD is an expansion in the inverse {\it bare\/} heavy quark mass,
all meson mass differences can be obtained from correlation functions at 
$\vec p = \vec 0$, but the absolute meson mass itself remains undetermined.
One way to fix the mass is to compute the change in energy when a meson is
boosted,
\begin{equation}\label{Mkin}
   E_{\bf p} - E_0 = \frac{\bf p^2}{2M_{\rm kin}}~.
\end{equation}
This defines the kinetic mass, $M_{\rm kin}$, which is interpreted as the
meson's physical mass.  For the present work, $E_{\bf p}$ is computed
only for the ${}^1S_0$ state, with ${\bf p} = (0,0,2\pi/L_s)$ where $L_s$ is 
the spatial extent of the lattice.

\section{RESULTS}

Gauge field configurations, periodic at all lattice boundaries, were 
generated using a pseudo-heatbath algorithm.
After 4000 thermalizing sweeps, the retained configurations were separated
from one another by 250 sweeps.
Light quark matrix inversion was performed by a stabilized biconjugate
gradient algorithm, also periodic at the lattice boundaries.  

\begin{figure}[tbh]
\epsfxsize=380pt \epsfbox[30 025 498 700]{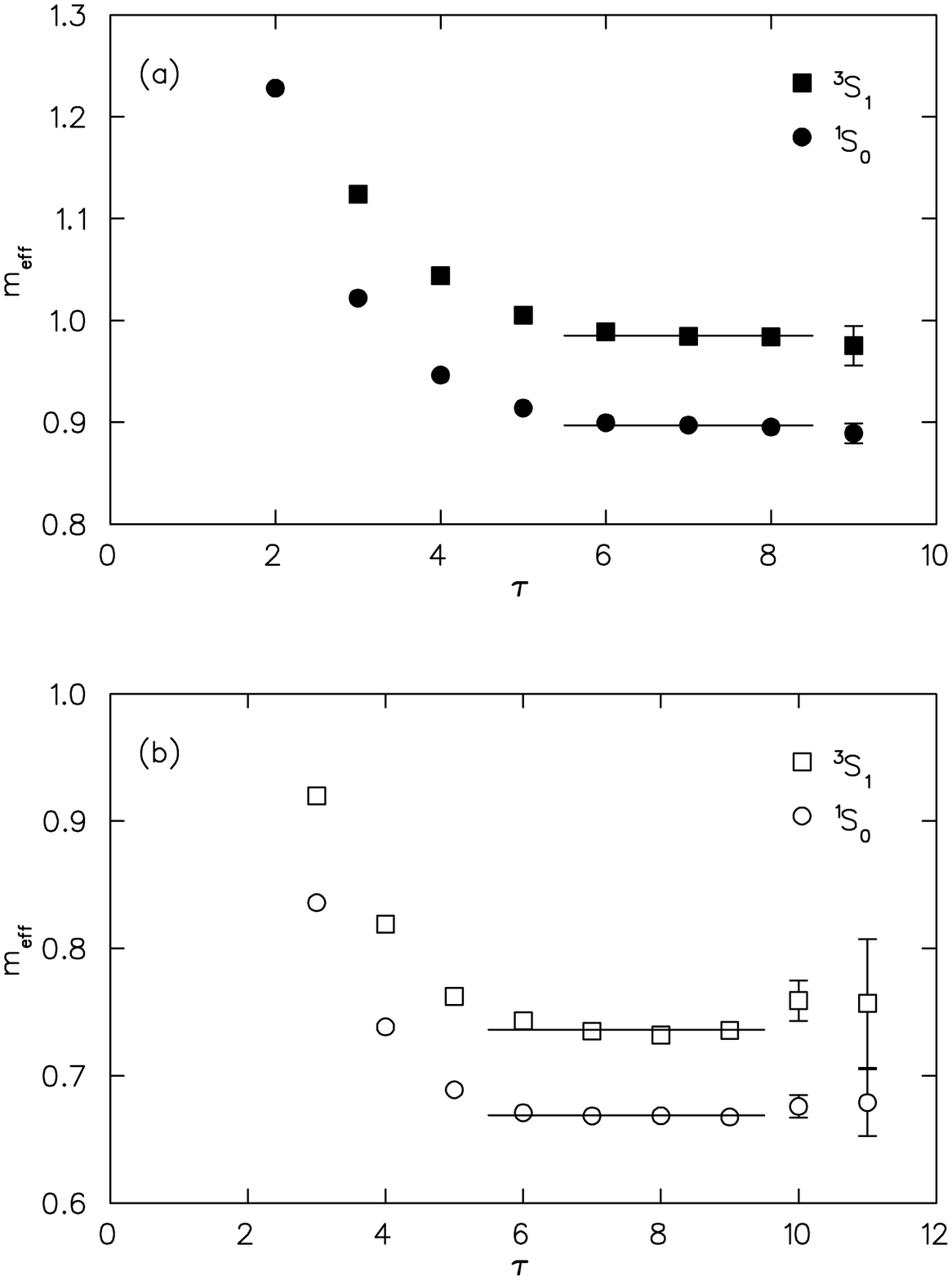}
\vspace{18pt}
\caption{Effective mass plots for ${}^1S_0$ and ${}^3S_1$ charmed mesons
         at rest, including terms up to $O(1/M^3)$.
         Solid symbols denote data at $\beta=6.8$, $aM=1.43$
         and $\kappa=0.135$, while open symbols correspond to $\beta=7.0$,
         $aM=1.10$ and $\kappa=0.134$.  The meson source is at $\tau=1$.
         Horizontal lines indicate masses extracted from the plateau regions.
         }\label{fig:effmass}
\end{figure}

Although the light quark field is periodic in each space-time direction,
Eqs.~(\ref{evol1}-\ref{evol2}) indicate that the heavy quark field is periodic 
only in the spatial directions.  Therefore, the correlation functions for
heavy-light mesons are only useful for times smaller than about $L_t/2$, where
$L_t$ is the temporal extent of the lattice.
Fig. \ref{fig:effmass} shows examples of effective mass plots for ${}^1S_0$
and ${}^3S_1$ charmed mesons, where
\begin{equation}
m_{\rm eff}(\tau) = -{\rm ln}\left(\frac{G_{\rm meson}(\vec{p},\tau+1)}
                              {G_{\rm meson}(\vec{p},\tau)}\right)~.
\end{equation}
In all cases, a plateau is found where the effective mass values computed at 
three (or more) neighboring timeslices are equal within the bootstrap errors.

The mass of a ${}^1S_0$ heavy-light meson is taken to be the average of
all effective mass values within the plateau region.  The uncertainty
associated with this mass is determined by a bootstrap procedure, where
1000 bootstrap ensembles are chosen from the original data such that
each ensemble is the same size as the original data sample.
A bootstrap distribution is then obtained by computing, for each ensemble,
the average of all effective mass values within the plateau region.
The uncertainty in the meson mass is then taken to be half the distance 
between the 16th and 84th percentiles in the bootstrap distribution.

The same method could be applied to the mass of the ${}^3S_1$, but the
errors are expected to be highly correlated with those in the 
${}^1S_0$ calculation.
Therefore, the ${}^3S_1 - {}^1S_0$ mass difference is determined by
applying the bootstrap technique directly to the mass difference.

Some parameters of the simulations are given in Table~\ref{tab:params}.
\begin{table*}
\caption{Simulation parameters.  $N_U$ is the number of gauge field
         configurations, $\kappa_c$ is the hopping parameter at the critical
         point, $\kappa_s$ is the hopping parameter corresponding
         to the strange quark mass (from $m_{K^*}/m_K$=1.8) and
         $a_\rho$ is the lattice spacing derived from the $\rho$ meson mass.
         }\label{tab:params}
\begin{tabular}{rccc|cccc}
Lattice & $N_U$ & $\beta$ & $\kappa$ & $U_0$ &
                      $\kappa_c$ & $\kappa_s$ & $a_\rho$[fm] \\
\hline
$8^3 \times 14$
& 400 & 6.8 & 0.135, 0.138, 0.141 & 0.854 & 0.1458(1) & 0.1398(4) & 0.260(6) \\
$10^3 \times 16$
& 300 & 7.0 & 0.134, 0.137, 0.140 & 0.865 & 0.1434(1) & 0.1385(3) & 0.225(8)
\end{tabular}
\end{table*}
The determinations of $\kappa_c$, $\kappa_s$ and $a_\rho$ come
from separate simulations
involving 250 configurations at $\beta=6.8$ and 200 configurations
at $\beta=7.0$.  The values of $U_0$ agree with Ref. \cite{Trottier}.
The stabilizing parameter $n$ of Eqs.~(\ref{evol1}-\ref{evol2}) was at least
as large as $n=4$ for $aM < 1.2$, $n=3$ for $1.2 \leq aM < 1.5$ and $n=2$ for 
$aM \geq 1.5$.\cite{Trottier}

The bare charm quark masses at $\beta=6.8$ and $\beta=7.0$
were obtained in Ref. \cite{Trottier} by equating
the kinetic mass of the $\eta_c$ with its physical mass.
This calculation can be reproduced (with poorer statistics) using the gauge 
field configurations of Table \ref{tab:params}, provided that one additional
term is added to the Hamiltonian of Eqs.~(\ref{H}-\ref{dH3}),
\begin{equation}
  H_{full} = H - \frac{c_8}{U_0^4}\frac{3g}{64M^4}
               \left\{\tilde\Delta^{(2)},\mbox{{\boldmath$\sigma$}}
               \cdot(\tilde{\bf \Delta}\times\tilde{\bf E}-\tilde{\bf E}\times 
               \tilde{\bf \Delta})\right\}~.
\end{equation}
For heavy-light calculations the term containing $c_8$ is suppressed by four 
powers of $1/M$, but in the velocity expansion relevant to quarkonium it 
contributes at $O(v^6)$, and
is the only $O(v^6)$ term absent in Eqs.~(\ref{H}-\ref{dH3}).
As was done in all other terms, the parameter $c_8$ is here set to its 
classical value of unity.

It is also worth noting that $H_{full}$ contains terms which are beyond 
$O(v^6)$ in the velocity expansion.  
The minimal Hamiltonian up to $O(v^6)$ is obtained from $H_{full}$ by omitting
the portions of the $c_9$ and $c_{10}$ terms which involve two powers of
$\bf\tilde{B}$.  It is typical to neglect the $c_{11}$ term as well.  In
Ref. \cite{Trottier}, the entire $c_{10}$ term was also omitted, since it
contains no spin structure and therefore seemed negligible for a discussion
of quarkonium spin splittings.  (This point will be addressed below in the
context of heavy-light mesons.)
The resulting Hamiltonian will be referred to as $H_{spin}$.

Table \ref{tab:M} shows a good agreement between the bare charm quark 
masses obtained here and those of Ref. \cite{Trottier}.
\begin{table*}
\caption{Values for the lattice spacing as derived 
        from the $1P$-$1S$ mass splitting of charmonium, and values for
        the heavy quark masses.
        Both masses from Ref. \protect\cite{Trottier} lead to 
        $M_{\eta_c}=2.9(1)$GeV.  All other entries show the range of $aM_c$ and 
        $aM_b$ which reproduce the actual experimental values
        for $M_{\eta_c}$ and $M_\Upsilon$.}\label{tab:M}
\begin{tabular}{c|c|ccc|cc}
$\beta$ & $a_{\rm hvy}$[fm] & 
     \multicolumn{3}{c}{$aM_c$} & \multicolumn{2}{c}{$aM_b$} \\
\cline{3-7}
 & (Ref. \protect\cite{Trottier}) & $H_{spin}$ (Ref. \protect\cite{Trottier}) 
 & $H_{spin}$ & $H_{full}$ & $H_{spin}$ & $H_{full}$ \\
\hline
6.8 & 0.257(9) & 1.43 & 1.5(1) & 1.7(1) & 5.0(2) & 5.0(2) \\
7.0 & 0.205(9) & 1.10 & 1.1(1) & 1.1(1) & 4.2(1) & 4.2(1) 
\end{tabular}
\end{table*}
The bare bottom quark mass can be obtained in a similar fashion, using the
$\eta_b$ in place of the $\eta_c$.  Because the $\eta_b$ has not yet been
seen experimentally, it's ``physical mass'' is obtained by subtracting
the hyperfine splitting (of the present lattice simulations) from the
experimental $\Upsilon$ mass.  In practice, the hyperfine splitting is a
negligible subtraction in comparison with the simulation uncertainties.
The resulting values for $M_b$ are shown in Table \ref{tab:M}.

Table \ref{tab:M} indicates that for $M \approx M_b$, the terms which 
distinguish
between $H_{spin}$ and $H_{full}$ are negligibly small in comparison with the
computational uncertainties.  Some evidence of their effect might be seen
near $M = M_c$, but the large uncertainties do not allow a definitive
statement to be made.  A more precise comparison of $H_{spin}$ and $H_{full}$
can be obtained from the quarkonium spin splittings, but this is not required
for the present study.

\begin{figure}[tbh]
\epsfxsize=380pt \epsfbox[30 025 498 700]{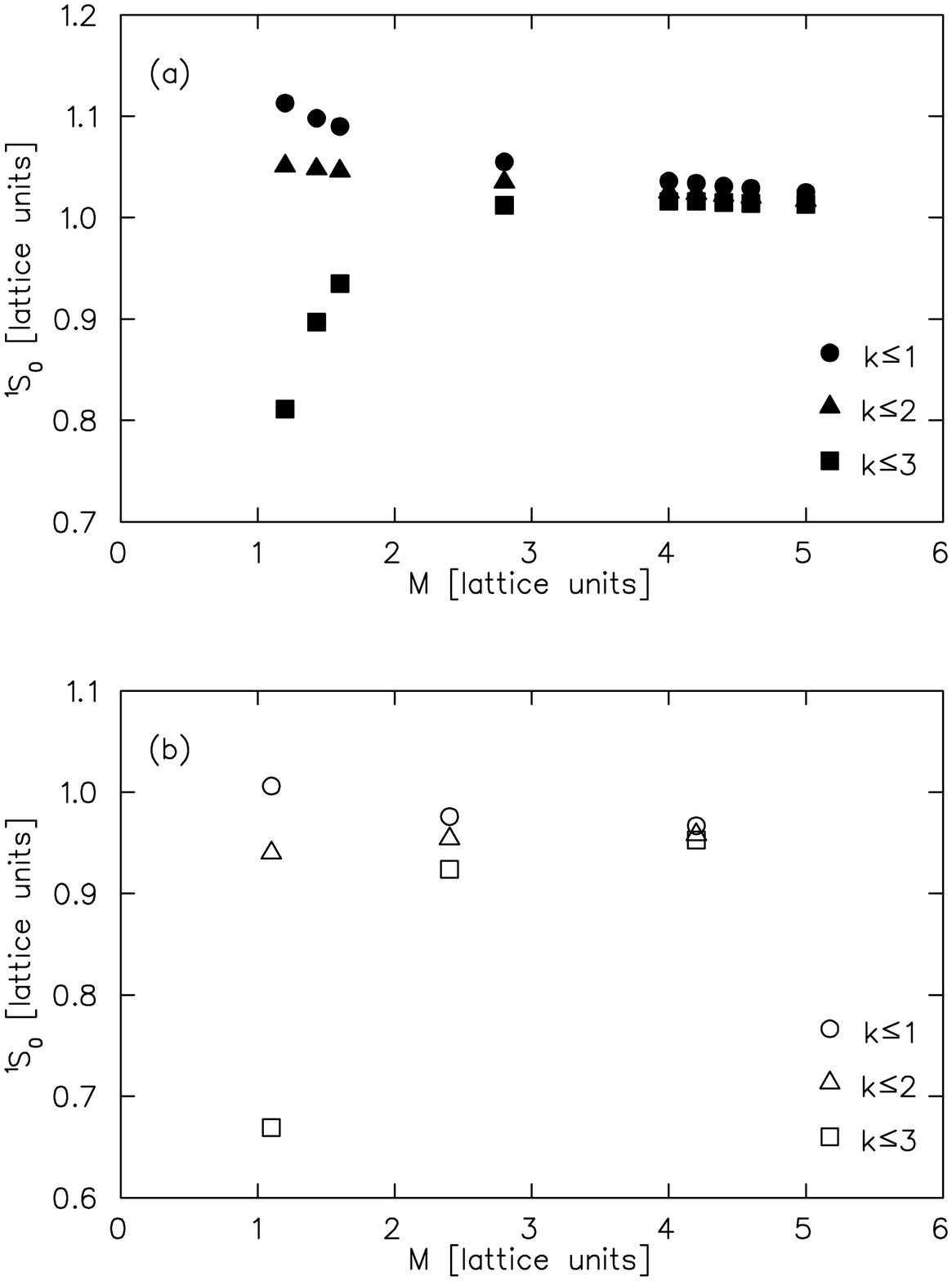}
\vspace{18pt}
\caption{The simulation energy of a ground state heavy-light meson at rest.
         Results are displayed from terms up to $O(1/M^k)$, with $k=1,2,3$.  
         $M$ is the bare heavy quark mass.
         Solid symbols denote data at $\beta=6.8$
         and $\kappa=0.135$, while open symbols correspond to $\beta=7.0$ and
         $\kappa=0.134$. 
         }\label{fig:EM}
\end{figure}

Having fixed all lattice parameters from light-light and heavy-heavy meson
observables, the heavy-light spectrum will now be considered.  
Fig. \ref{fig:EM} shows the simulation energy of a ${}^1S_0$ heavy-light
meson as a function of the bare heavy quark mass,
computed to $O(1/M)$, $O(1/M^2)$ and $O(1/M^3)$.
For each $\beta$, the light quark mass is fixed at a value slightly less
than twice the strange quark mass, according to Table \ref{tab:params}.  
Fig. \ref{fig:EM} indicates that terms beyond $O(1/M)$ provide small
corrections to the leading order result when $M \approx M_b$, but these
corrections grow as $M$ decreases.  Near $M = M_c$, the effect of $O(1/M^3)$
terms is larger than the $O(1/M^2)$ terms.

\begin{figure}[tbh]
\epsfxsize=380pt \epsfbox[30 419 498 732]{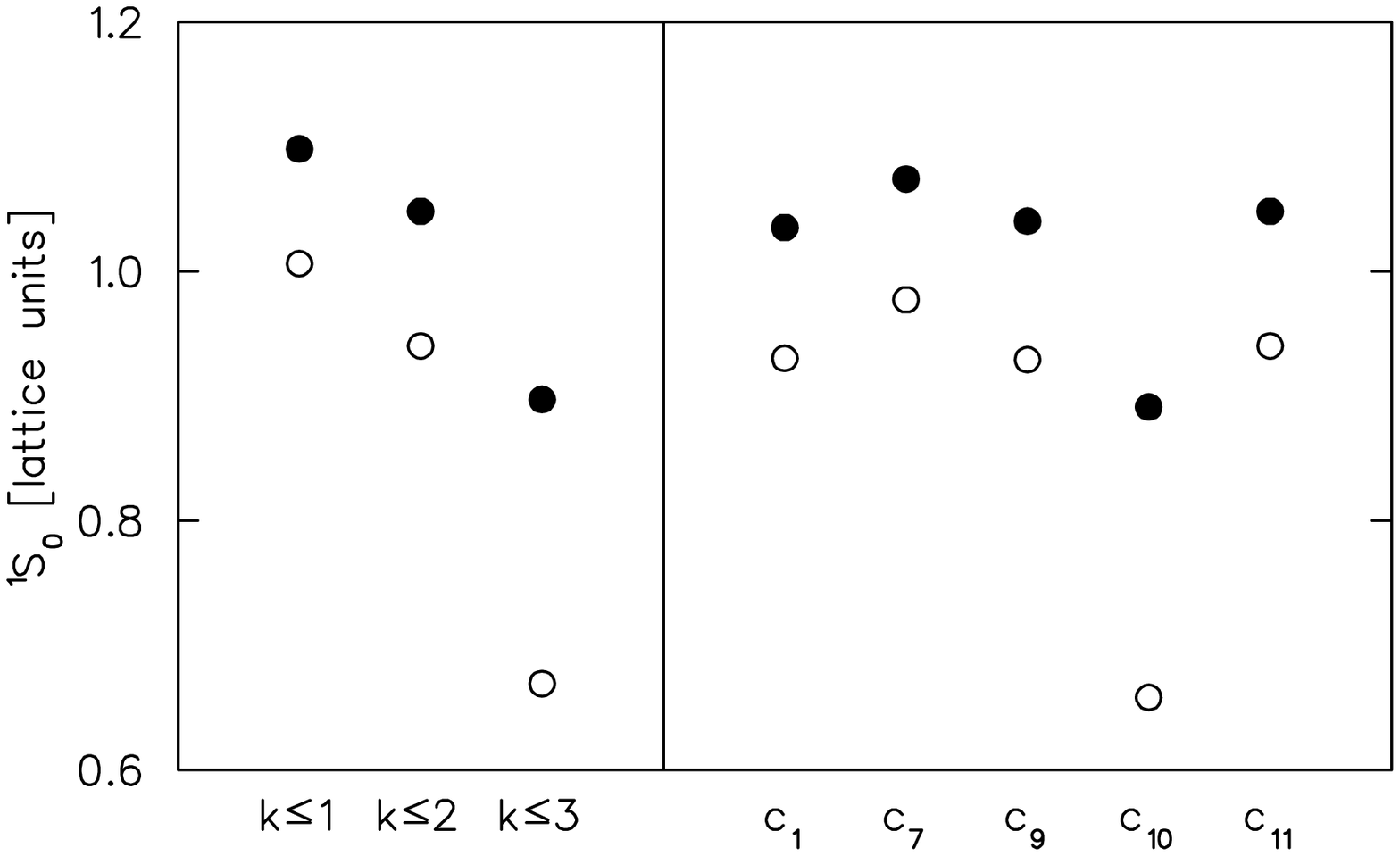}
\vspace{18pt}
\caption{The simulation energy of a ground state charmed meson at rest.
         Results are displayed from terms up to 
         $O(1/M^k)$, with $k=1,2,3$.  Solid symbols denote data at $\beta=6.8$,
         $\kappa=0.135$ and $aM=1.43$, while open symbols correspond to 
         $\beta=7.0$, $\kappa=0.134$ and $aM=1.10$.  To the right of the 
         vertical line, the effect of adding each $O(1/M^3)$ term to the 
         $O(1/M^2)$ Hamiltonian is shown individually.
         }\label{fig:Eord3}
\end{figure}

In order to understand the origin of such large $O(1/M^3)$ contributions
in the charm region, simulations were performed with each $O(1/M^3)$
term added individually to the lower-order Hamiltonian.
Results for the simulation energy are shown
in Fig. \ref{fig:Eord3}.  Apparently, all terms except the $c_{10}$ term 
offer only modest corrections to the lower-order result.
In fact, the $c_{10}$ term is unique because it is the only term in the
Hamiltonian (up to $O(1/M^3)$) which has a nonzero vacuum expectation 
value.\cite{Lepage}
This vacuum value is simply an $O(1/M^3)$ shift of the bare heavy quark mass,  
and therefore produces an $O(1/M^3)$ shift in the meson simulation energy.

\begin{figure}[tbh]
\epsfxsize=380pt \epsfbox[30 419 498 732]{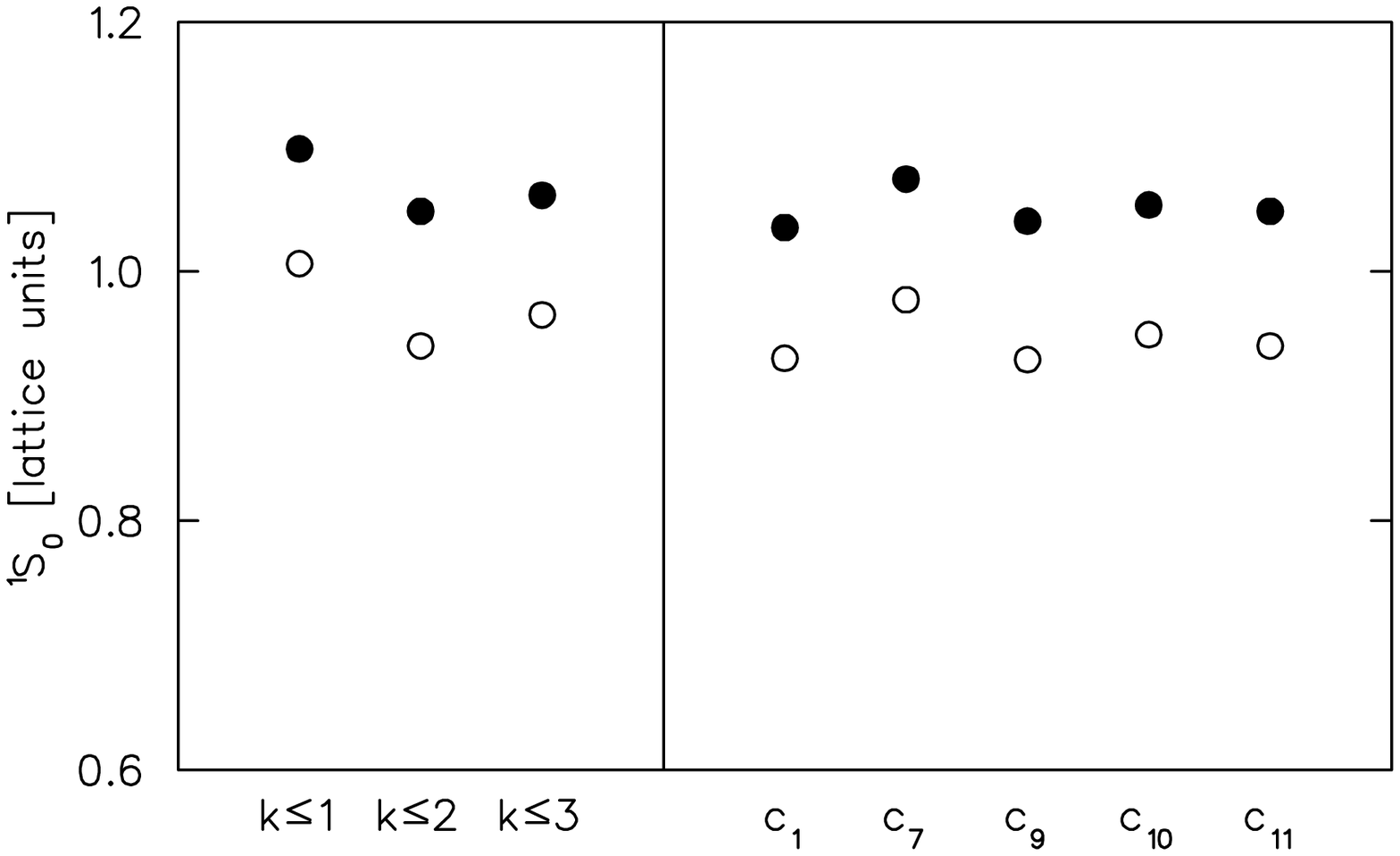}
\vspace{18pt}
\caption{These data are identical to Fig. \protect\ref{fig:Eord3} except that
         the vacuum expectation value
         has here been subtracted from the $c_{10}$ term.
         }\label{fig:Eord3GG}
\end{figure}

The vacuum expectation value can be computed directly from the gauge field
configurations of Table \ref{tab:params} by averaging over all lattice sites,
\begin{equation}\label{GG}
a^4\left<\frac{g^2}{U_0^8}(\tilde{\bf E}^2+\tilde{\bf B}^2)\right>
    = \left\{\begin{tabular}{l}
                  13.6161(20), for $\beta=6.8$ \\
                  12.0716(16), for $\beta=7.0$~.
             \end{tabular}\right.
\end{equation}
If the vacuum expectation value is removed from the action,
\begin{eqnarray}
   \delta{S}_Q &=& \frac{c_{10}}{8M^3}\int{\rm d}^4x\,{Q^a}^\dagger(x)
                   \frac{g^2}{U_0^8}(\tilde{\bf E}^2+\tilde{\bf B}^2)^{ab}
                   Q^b(x) \nonumber \\
   & \rightarrow & \frac{c_{10}}{8M^3}\int{\rm d}^4x\,{Q^a}^\dagger(x)\left[
                   \frac{g^2}{U_0^8}(\tilde{\bf E}^2+\tilde{\bf B}^2)^{ab}
                   -\frac{\delta^{ab}}{3}\left<\frac{g^2}{U_0^8}
                   (\tilde{\bf E}^2+\tilde{\bf B}^2)\right>\right]Q^b(x)~,
\end{eqnarray}
then the contribution of the $c_{10}$ term to the simulation energy is
reduced to the size of the other $O(1/M^3)$ terms, as shown in 
Fig. \ref{fig:Eord3GG}.
Removal of the vacuum expectation value from the action is justified 
because it is simply an addition to the heavy quark mass, which
has already been removed from the Hamiltonian, Eq.~(\ref{H}),
by the standard heavy-field transformation of HQET.
Mass {\it differences} should not depend on whether or not the vacuum 
expectation value remains explicitly in the Hamiltonian.

\begin{figure}[tbh]
\epsfxsize=380pt \epsfbox[30 025 498 700]{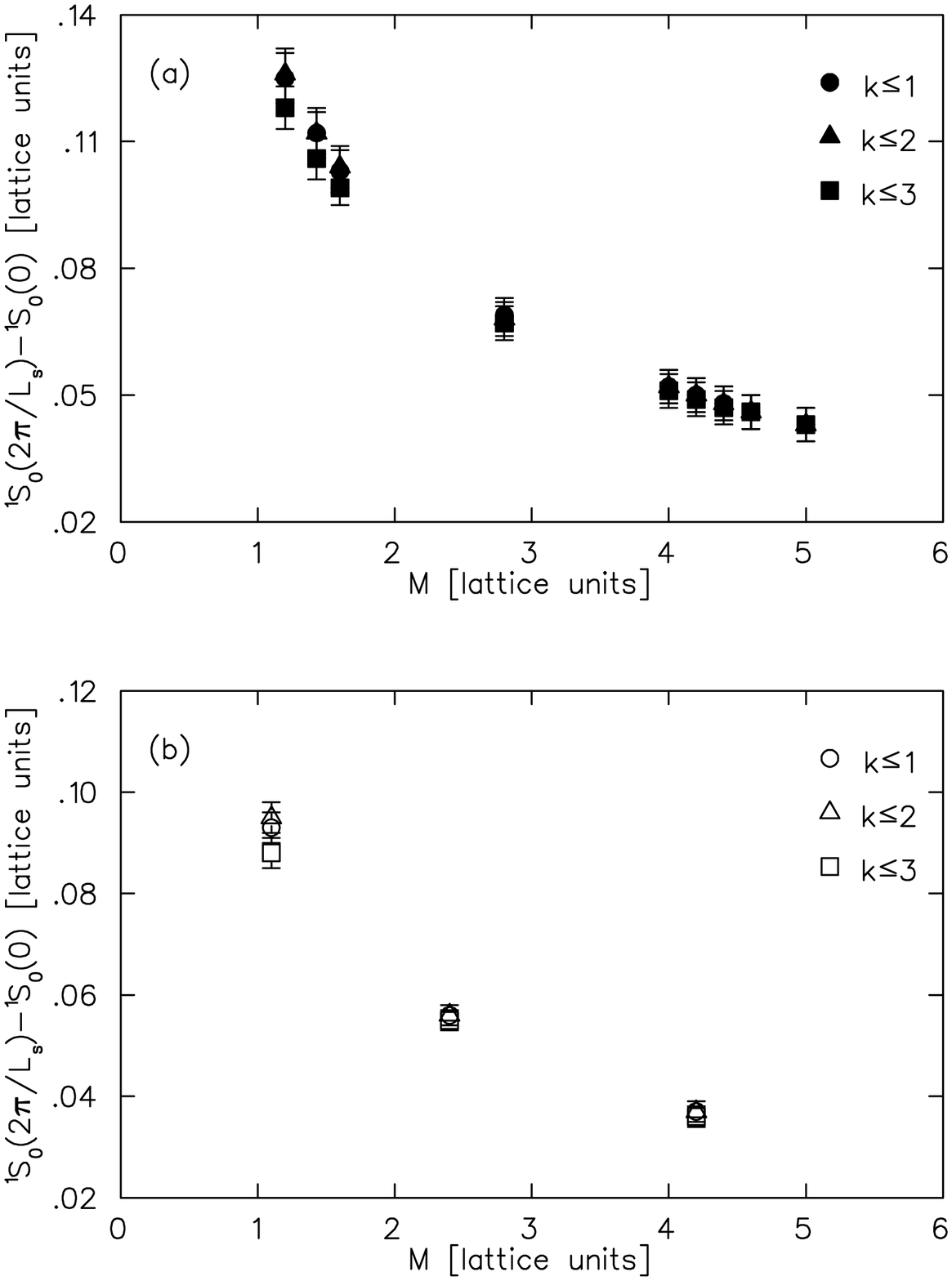}
\vspace{18pt}
\caption{The energy splitting between a ground state heavy-light meson with
         momentum $\vec p = (0,0,2\pi/L_s)$ (where $L_s$ is the spatial 
         extent of the lattice) and the same meson at rest.  
         Results are displayed from terms up to $O(1/M^k)$, with $k=1,2,3$.  
         $M$ is the bare heavy quark mass.
         Solid symbols denote data at $\beta=6.8$ 
         and $\kappa=0.135$, while open symbols correspond to $\beta=7.0$ and
         $\kappa=0.134$.  
         }\label{fig:kinM}
\end{figure}

Fig. \ref{fig:kinM} displays the energy splitting between a ${}^1S_0$ meson
at zero and nonzero 3-momenta.  The vacuum expectation value has not been
removed from the $c_{10}$ term, but its effect should cancel in the
difference between zero and nonzero 3-momenta.  Fig. \ref{fig:kinM} shows
that the $O(1/M^2)$ and $O(1/M^3)$
terms offer only small corrections to the leading contribution, so a
determination of the kinetic mass from Eq.~(\ref{Mkin}) will not depend
sensitively on the presence of these terms. 

\begin{figure}[tbh]
\epsfxsize=380pt \epsfbox[30 419 498 732]{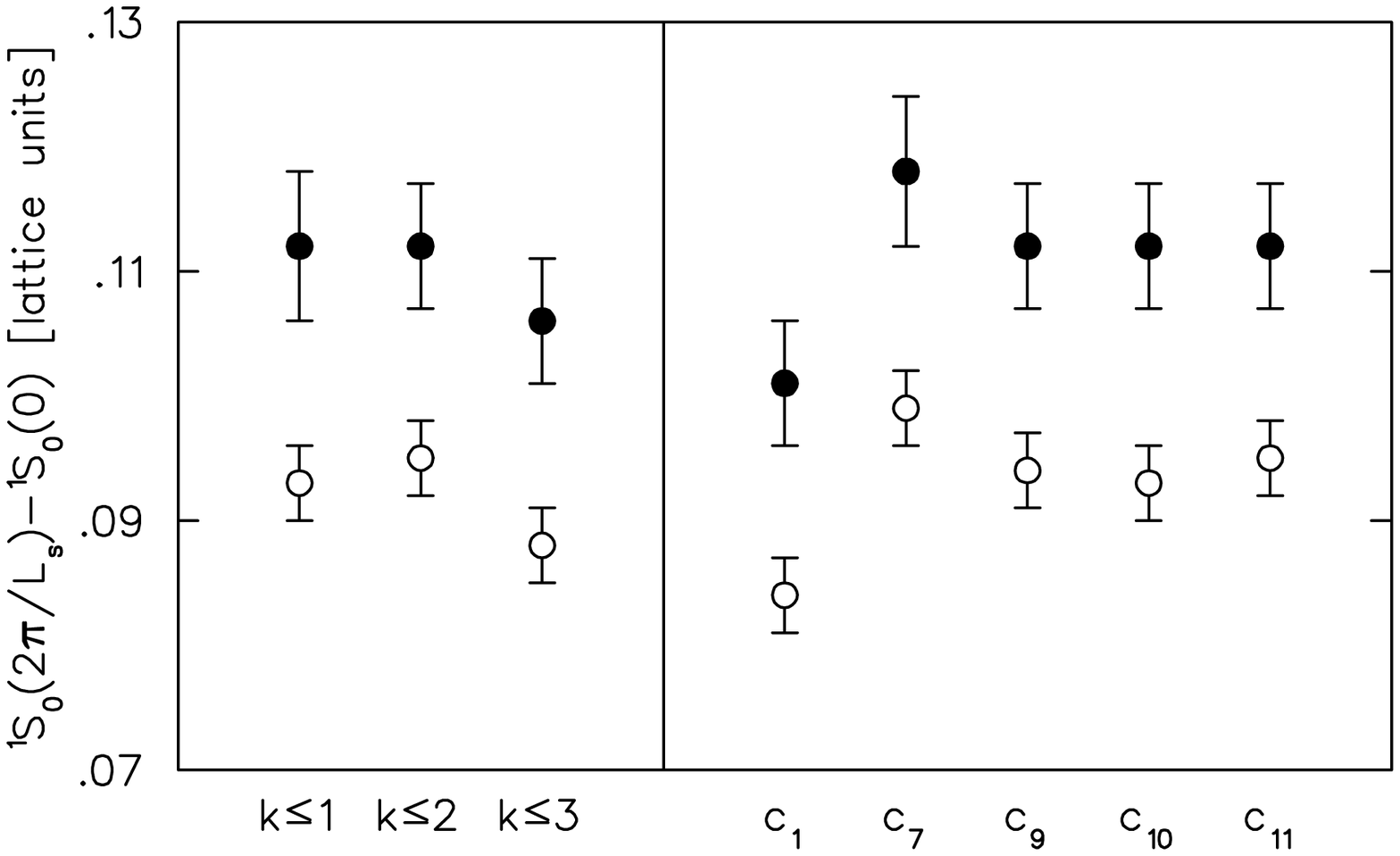}
\vspace{18pt}
\caption{The energy splitting between a ground state charmed meson with
         momentum $2\pi/L_s$ (where $L_s$ is the spatial extent of the lattice)
         and the same meson at rest.  Results are displayed from terms up to 
         $O(1/M^k)$, with $k=1,2,3$.  Solid symbols denote data at $\beta=6.8$,
         $\kappa=0.135$ and $aM=1.43$, while open symbols correspond to 
         $\beta=7.0$, $\kappa=0.134$ and $aM=1.10$.  To the right of the 
         vertical line, the effect of adding each $O(1/M^3)$ term to the 
         $O(1/M^2)$ Hamiltonian is shown individually.
         }\label{fig:kinord3}
\end{figure}

Fig. \ref{fig:kinord3} gives the contribution of each $O(1/M^3)$
term to the energy splitting between a ${}^1S_0$ meson at zero and
nonzero 3-momenta.  The contribution of each term is small in
comparison to the statistical uncertainties, including the $c_{10}$ term.
(As expected, both Figs. \ref{fig:kinM} and \ref{fig:kinord3} remain 
essentially unaltered if the vacuum expectation value is subtracted.)

Fig. \ref{fig:kinM} raises another important issue.
Because the bare quark mass is not a physical parameter,
it is not necessarily sensible to compare the results
of the $O(1/M)$, $O(1/M^2)$ and $O(1/M^3)$ Hamiltonians
at a fixed bare mass.  It would be preferable to
make the comparison at a fixed kinetic mass.
However, Fig. \ref{fig:kinM} demonstrates that the $O(1/M)$,
$O(1/M^2)$ and $O(1/M^3)$ Hamiltonians all give rise to the same 
relationship between the bare and kinetic masses, within the statistical 
uncertainties.

\begin{figure}[tbh]
\epsfxsize=380pt \epsfbox[30 025 498 700]{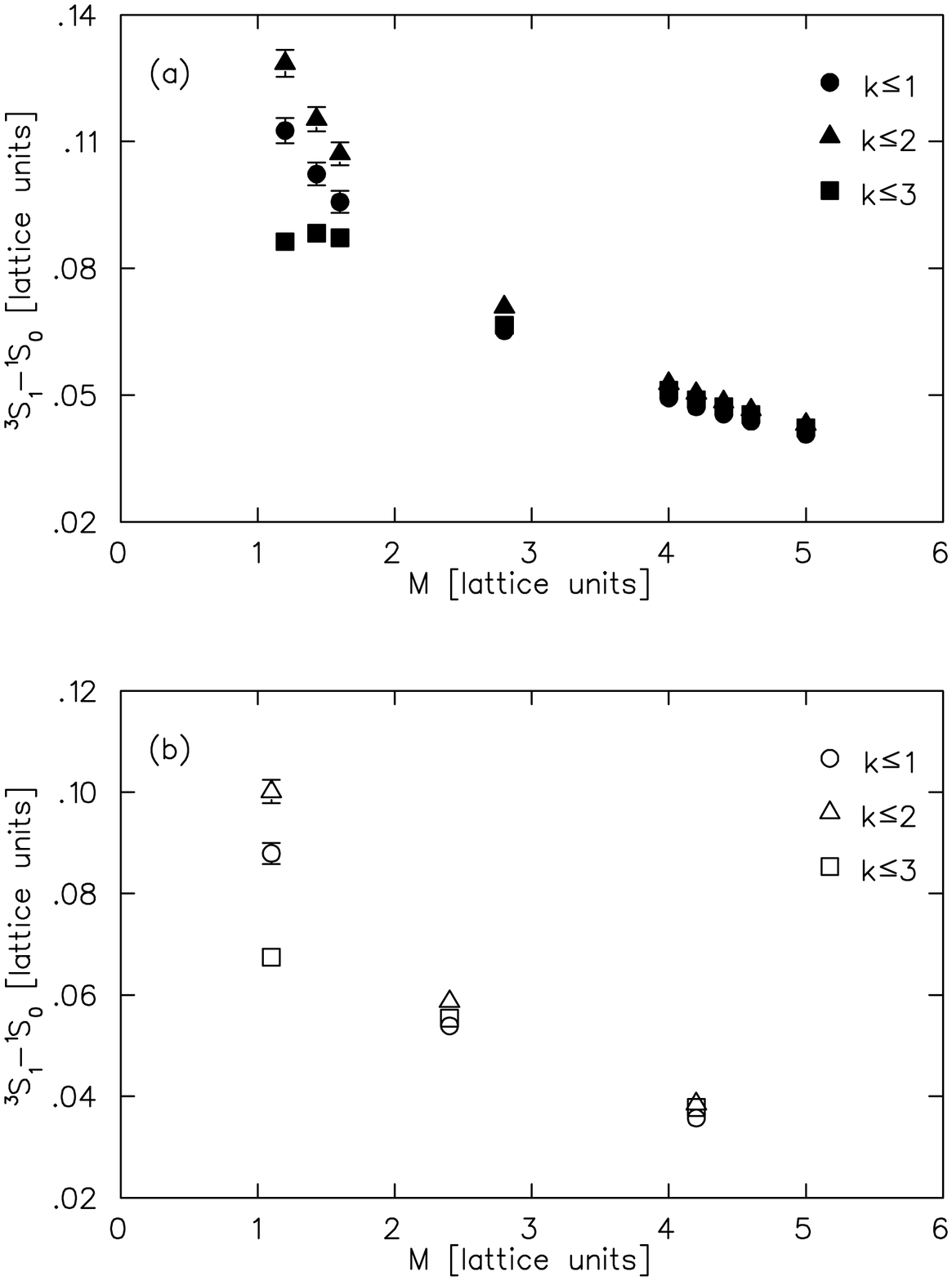}
\vspace{18pt}
\caption{The spin splitting of S-wave
         heavy-light mesons, from terms up to $O(1/M^k)$, with $k=1,2,3$.  
         $M$ is the bare heavy quark mass.
         Solid symbols denote data at $\beta=6.8$ and $\kappa=0.135$, while 
         open symbols correspond to $\beta=7.0$ and $\kappa=0.134$. 
         The vacuum expectation value has {\bf not} been removed
         from the $c_{10}$ term.
         }\label{fig:hypM}
\end{figure}

In Fig. \ref{fig:hypM}, the spin splitting is plotted as a function of the 
bare heavy quark mass.  The vacuum expectation value has not been removed
from the $c_{10}$ term.  $O(1/M^2)$ and $O(1/M^3)$ terms provide small 
corrections to the spin splitting near $M = M_b$, but sizeable ones near
$M = M_c$, with an $O(1/M^3)$ contribution that is larger in magnitude
than the $O(1/M^2)$ contribution. 

\begin{figure}[tbh]
\epsfxsize=380pt \epsfbox[30 419 498 732]{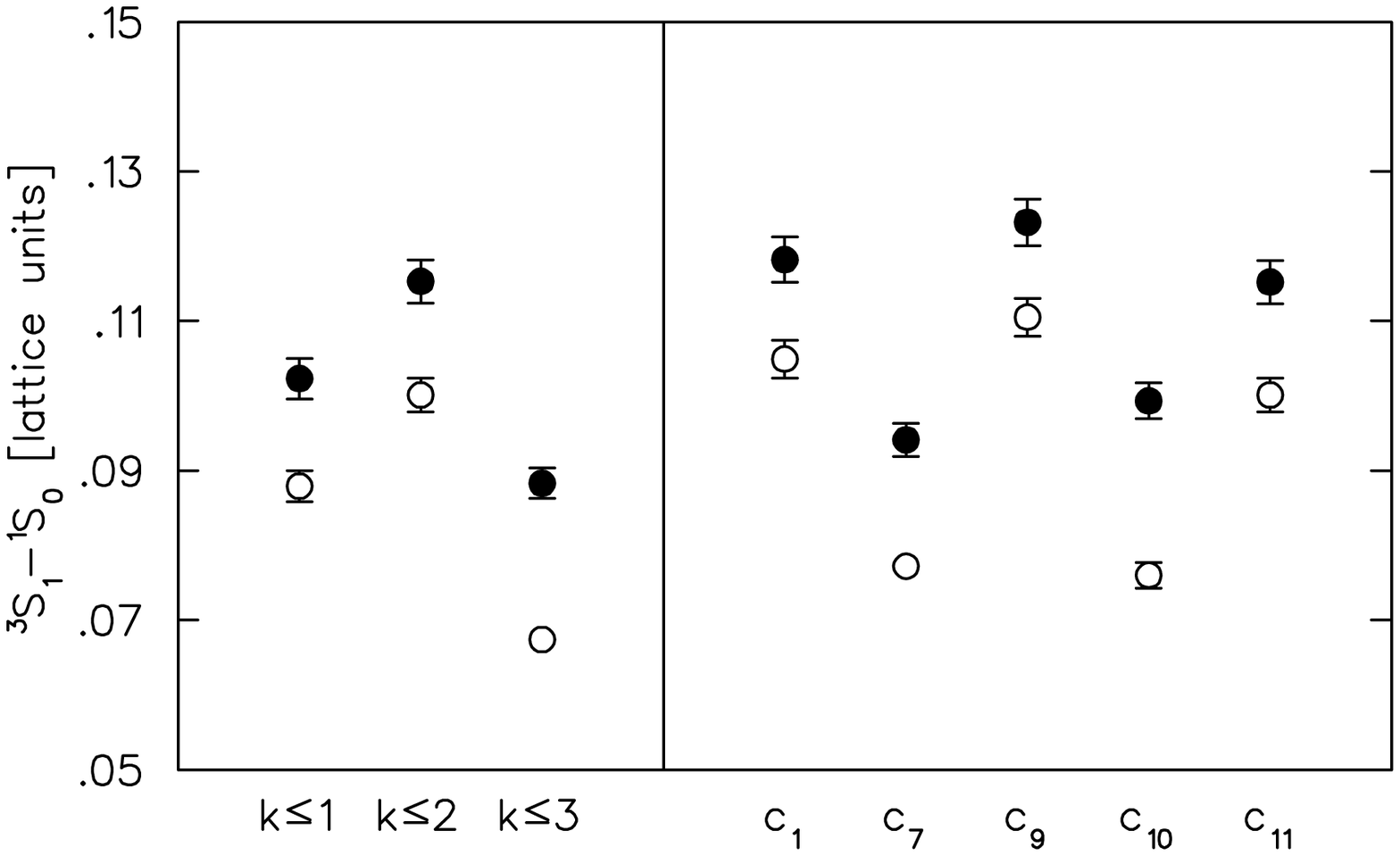}
\vspace{18pt}
\caption{The spin splitting of S-wave charmed mesons.  
         Results are displayed from terms up to 
         $O(1/M^k)$, with $k=1,2,3$.  Solid symbols denote data at $\beta=6.8$,
         $\kappa=0.135$ and $aM=1.43$, while open symbols correspond to 
         $\beta=7.0$, $\kappa=0.134$ and $aM=1.10$.  To the right of the 
         vertical line, the effect of adding each $O(1/M^3)$ term to the 
         $O(1/M^2)$ Hamiltonian is shown individually.
         The vacuum expectation value has {\bf not} been removed
         from the $c_{10}$ term.
         }\label{fig:hypord3}
\end{figure}
\begin{figure}[tbh]
\epsfxsize=380pt \epsfbox[30 419 498 732]{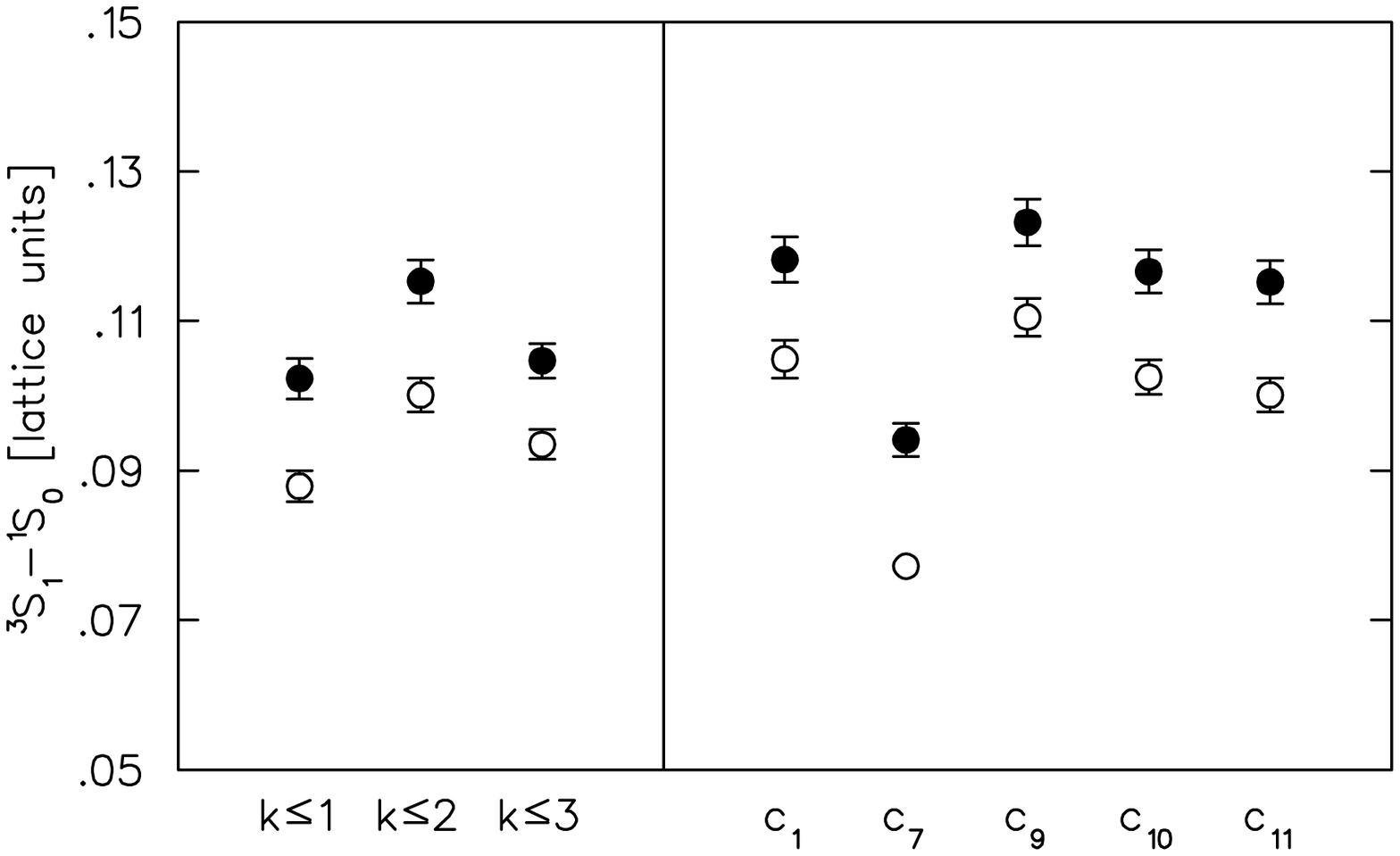}
\vspace{18pt}
\caption{These data are identical to Fig. \protect\ref{fig:hypord3} except that
         the vacuum expectation value
         has here been subtracted from the $c_{10}$ term.
         }\label{fig:hypord3GG}
\end{figure}

According to Fig. \ref{fig:hypord3}, there are two $O(1/M^3)$ terms which
dominate the large correction to the spin splitting: $c_7$ and $c_{10}$.
The importance of $c_{10}$ for the spin splitting is somewhat puzzling,
since that term in the Hamiltonian is spin-independent.
Some insight is gained by removing the vacuum expectation value from the
$c_{10}$ term\cite{Lepage}, which is shown in Fig. \ref{fig:hypord3GG} to 
remove almost the entire effect of the $c_{10}$ term in the charm region.
This small contribution of $c_{10}$ to the spin splitting is to be expected
for a spin-independent operator.  
Apparently the large effect of $c_{10}$ in Fig. \ref{fig:hypord3}
is spurious, perhaps because the vacuum expectation value is not sufficently
small compared to the heavy quark mass.

To confirm this, recall Eq.~(\ref{evol2}) which describes the discretization
of the heavy quark propagator.  In the continuum limit, this propagation 
depends exponentially on the Hamiltonian.
At a finite lattice spacing, Eq.~(\ref{evol2}) uses a linear
approximation to the exponential of $a\delta{H}$.  Following
Ref. \cite{NRQCD2}, the exponential of $aH_0$ was fit more precisely
than a simple linear approximation,
so that an instability could be avoided at small $M$.
In Eq.~(\ref{evol2}), if the $c_{10}$ term is subtracted from $\delta{H}$ and
added to $H_0$, then the spin splitting reproduces Fig. \ref{fig:hypord3GG}
rather than Fig. \ref{fig:hypord3}.
(The values of the parameter $n$ in Eq.~(\ref{evol2}) are not changed from
what have been used througout this work.  For example, 
$n=3$ at $\beta=6.8$ and $n=4$ at $\beta=7.0$ in the charm region.)
This supports the suspicion that the large vacuum expectation value was 
causing a breakdown of the results in Fig. \ref{fig:hypord3}.  The problem
has thus been successfully overcome in two separate ways: 
by the explicit removal of the vacuum value
from the Hamiltonian, or by a better-than-linear approximation to the
exponential dependence of heavy quark propagation on the $c_{10}$ term.

\begin{figure}[tbh]
\epsfxsize=380pt \epsfbox[30 025 498 700]{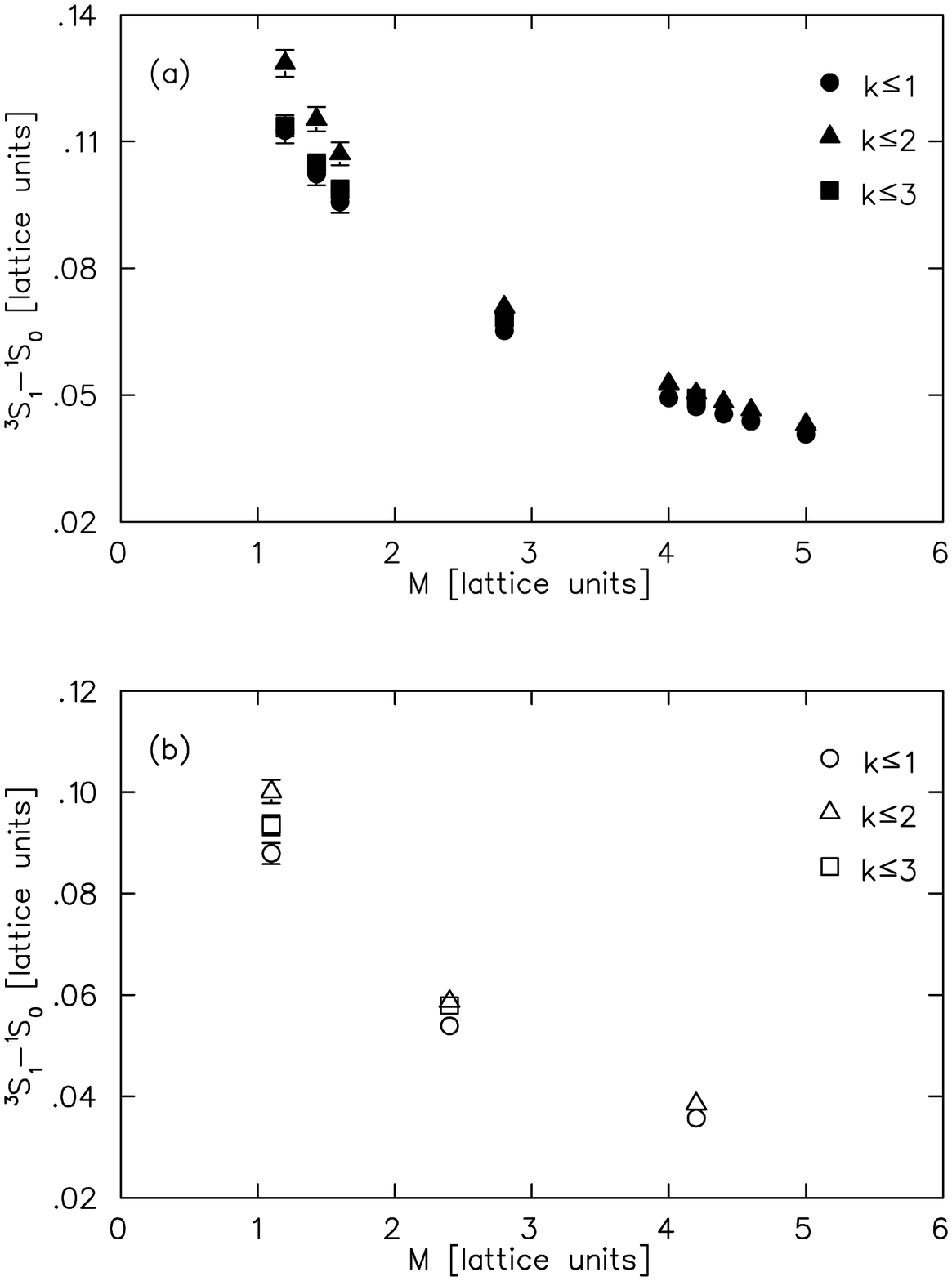}
\vspace{18pt}
\caption{These data are identical to Fig. \protect\ref{fig:hypM} except that
         the vacuum expectation value
         has here been subtracted from the $c_{10}$ term.
         }\label{fig:hypMGG}
\end{figure}

Fig. \ref{fig:hypord3GG} indicates that the spin splitting satisfies
$|O(1/M^3)| < |O(1/M^2)| < |O(1/M)|$.
This same ordering persists for all $M$ values considered, as seen in
Fig. \ref{fig:hypMGG}.  
Notice in particular that the turnover of the $O(1/M^3)$ data near $M_c$ is 
completely removed from Fig. \ref{fig:hypM} by correctly accounting for the
vacuum value in the $c_{10}$ term.  
Meanwhile, the data at $aM > 2$ are 
not affected in a statistically-significant way.
The $1/M$ expansion might now appear to be nicely convergent,
but some caution is suggested due to the $c_7$ term, which is by itself
larger in magnitude than the total $O(1/M^2)$ contribution to the 
spin splitting of a charmed meson.

It is interesting to consider the difference between squares of the ${}^3S_1$
and ${}^1S_0$ masses, which empirically is remarkably independent of the 
``heavy'' quark mass:
\begin{eqnarray}\label{diffsqexp1}
   (B^{+*})^2-(B^+)^2 \approx (B^{0*})^2-(B^0)^2 &=& 0.48~{\rm GeV}^2 \\
   (B_s^*)^2-(B_s)^2 &=& 0.51~{\rm GeV}^2 \\
   (D^{+*})^2-(D^+)^2 \approx (D^{0*})^2-(D^0)^2 &=& 0.55~{\rm GeV}^2 \\
   (D_s^*)^2-(D_s)^2 &=& 0.59~{\rm GeV}^2 \\
   (K^{+*})^2-(K^+)^2 \approx (K^{0*})^2-(K^0)^2 &=& 0.56~{\rm GeV}^2 \\
   (\rho^+)^2-(\pi^+)^2 \approx (\rho^0)^2-(\pi^0)^2 &=& 0.57~{\rm GeV}^2
   \label{diffsqexp2}
\end{eqnarray}
In the extreme heavy quark limit, this result is easily understood using
heavy quark symmetry:
the spin splitting vanishes as $1/M$ while the meson masses themselves
grow linearly with $M$, so the difference of squares is a constant,
\begin{equation}\label{sqdiff}
   m_V^2 - m_P^2 = (m_V-m_P)(m_V+m_P) = {\rm constant} + O(1/M)~.
\end{equation}
For mesons containing only light quarks, the explanation is perhaps not so
clear.  Some authors have related it to chiral symmetry\cite{ChiSym}.
It should be noted that $m_V^2 - m_P^2 \approx constant$ is also a 
consequence of the nonrelativistic quark model with a linear potential 
(and no heavy quark assumptions).

\begin{figure}[tbh]
\epsfxsize=380pt \epsfbox[30 025 498 700]{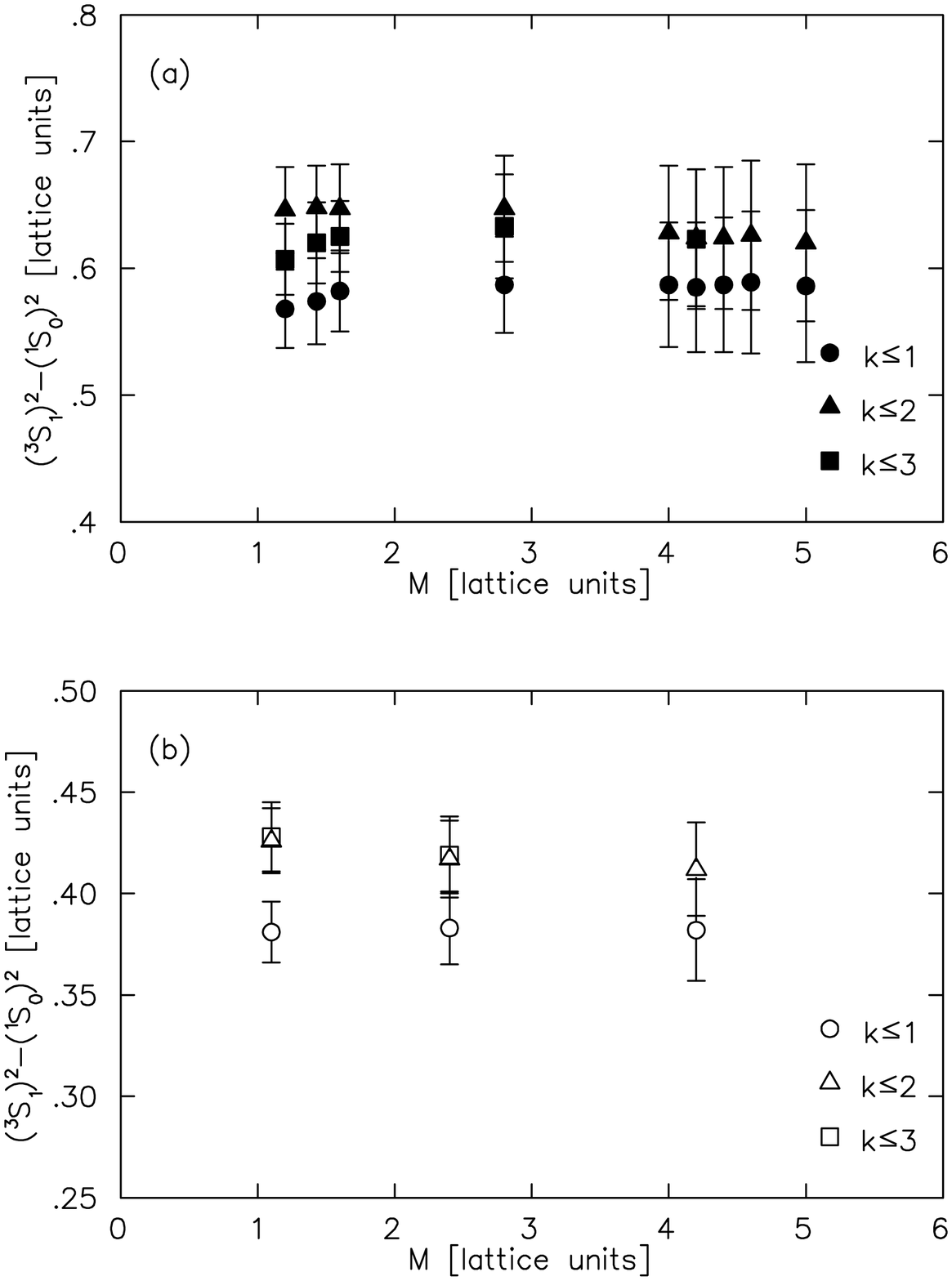}
\vspace{18pt}
\caption{The difference between squared masses of vector and pseudoscalar
         heavy-light mesons, from terms up to $O(1/M^k)$, with $k=1,2,3$.  
         $M$ is the bare heavy quark mass.
         Solid symbols denote data at $\beta=6.8$ and $\kappa=0.135$, while 
         open symbols correspond to $\beta=7.0$ and $\kappa=0.134$. 
         The vacuum expectation value has been subtracted from the 
         $c_{10}$ term.
         }\label{fig:VVPPMGG}
\end{figure}

The difference of squares arising from the present simulations is shown 
in Fig. \ref{fig:VVPPMGG} with the vacuum value subtracted from the 
$c_{10}$ term, and for a particular light quark $\kappa$.  (For all
cases considered, the results are essentially independent of $\kappa$, in
agreement with Eqs.~(\ref{diffsqexp1}-\ref{diffsqexp2}).)  The large
errors are due to the required use of $M_{\rm kin}$ and they are 
correlated, as evidenced by the central values being constant to within a
much smaller uncertainty than the quoted errors would require.
In fact, the $O(1/M)$, $O(1/M^2)$ and $O(1/M^3)$ data are each constant 
for the full range of $M$-values that were considered.

\begin{figure}[tbh]
\epsfxsize=380pt \epsfbox[30 025 498 700]{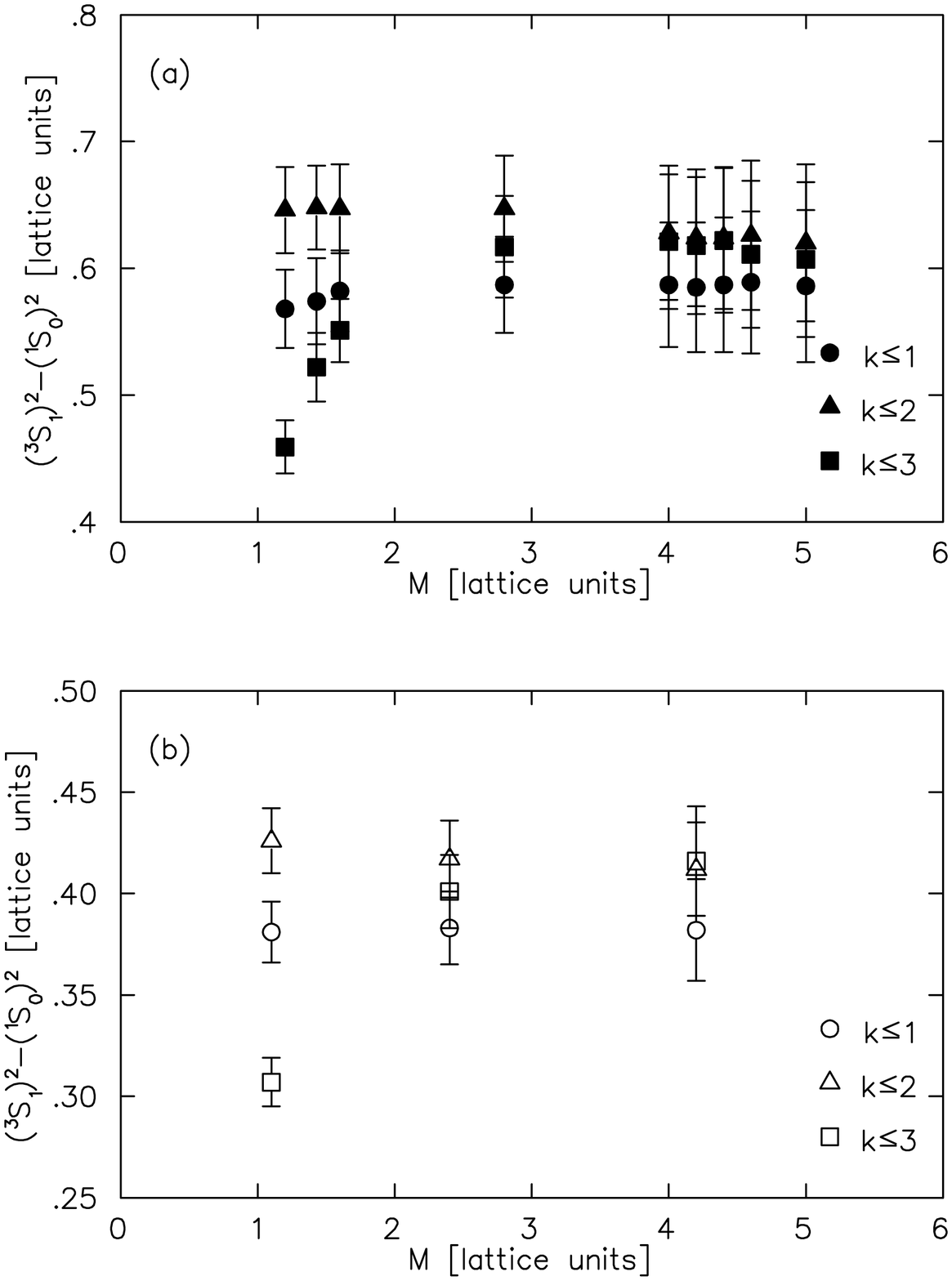}
\vspace{18pt}
\caption{These data are identical to Fig. \protect\ref{fig:VVPPMGG} except that
         the vacuum expectation value
         has here {\it not\/} been subtracted from the $c_{10}$ term.
         }\label{fig:VVPPM}
\end{figure}

For the purpose of comparison, the same plot
is displayed in Fig. \ref{fig:VVPPM},
but without a proper treatment of the vacuum expectation value, i.e.
with the vacuum value retained in the $c_{10}$ term and 
the simple heavy quark propagation of Eq.~(\protect\ref{evol2}).
Erroneous results are clearly produced at $O(1/M^3)$ for $aM<2$.

For all of the observables under discussion in this work, the $c_{11}$ term 
is essentially
irrelevant, which may not be too surprising in light of its large numerical
suppression factor in the Hamiltonian, Eq.~(\ref{dH3}).

Finally, to make the connection to experiment, it is necessary to interpolate 
to the strange quark mass (i.e. to $\kappa_s$), and to extrapolate to the 
limit of massless up and down quarks ($\kappa_c$).  
Interpolations are performed linearly between the two nearest
$\kappa$ values, and extrapolations are linear in all three available $\kappa$
values.  It is also necessary to determine the physical mass scale.
For $\beta=6.8$, both the $\rho$ meson mass and the charmonium $1P-1S$
mass splitting lead to the same physical scale ($a_\rho \approx a_{\rm hvy}$).
Use of the bare heavy quark masses from quarkonium (Table \ref{tab:M}) 
gives the results of Table \ref{tab:phys}.
\begin{table*}
\caption{Values for the physical masses and mass differences in MeV.
         The bare quark masses are set to $aM_c=1.43$, $aM_b=5.0$ at 
         $\beta=6.8$, and $aM_c=1.10$, $aM_b=4.2$ at $\beta=7.0$.  
         The Hamiltonian contains terms up to $O(1/M^k)$, where $k=1,2,3$.
         For charmed mesons, the vacuum value has been subtracted from the 
         $c_{10}$ term.  The quoted errors include
         the uncertainties in $\kappa_c$, $\kappa_s$ and $a$.  
         (The uncertainties in $\kappa_c$ and $\kappa_s$ are negligible except 
         for $D_s-D$ and $B_s-B$.)}\label{tab:phys}
\begin{tabular}{c|cccc|cccc}
    & $D_s$ & $D_s-D$ & $D^*-D$ & $D_s^*-D_s$
    & $B_s$ & $B_s-B$ & $B^*-B$ & $B_s^*-B_s$ \\
\hline
\multicolumn{9}{c}{$\bullet$ experiment (Ref. \protect\cite{PDG}) $\bullet$} \\
           & 1969 & 99,104 & 141,142 & 144
           & 5369(2) & 90(3) & 46 & 47(4) \\
\hline
\multicolumn{9}{c}{$\bullet$ $\beta=6.8$ $\bullet$} \\
$k \leq 1$ & 2010(130) & $96^{+6}_{-10}$ & 92(6) & 84(3)
           & 5300(700) & $75^{+9}_{-5}$ & 37(4) & 34(2) \\
$k \leq 2$ & 2000(110) & $99^{+6}_{-13}$ & 104(7) & 95(4)
           & 5300(700) & $79^{+10}_{-9}$ & 39(4) & 36(2) \\
$k \leq 3$ & 2090(120) & $98^{+6}_{-7}$ & 100(6) & 89(3)
           & 5300(700) & $79^{+8}_{-11}$ & 39(4) & 35(2) \\
\hline
\multicolumn{9}{c}{$\bullet$ $\beta=7.0$, scaled by $a_{\rm hvy}$ $\bullet$} \\
$k \leq 1$ & 1940(120) & $111^{+7}_{-11}$ & 100(10) & 92(5)
           & 5000(450) & $86^{+8}_{-7}$ & 40(4) & 37(2) \\
$k \leq 2$ & 1920(110) & $113^{+7}_{-11}$ & 114(8) & 105(5)
           & 5000(450) & $86^{+8}_{-7}$ & 43(4) & 40(2) \\
$k \leq 3$ & 2060(110) & $113^{+8}_{-10}$ & 111(7) & 100(5)
           & 5000(450) & $86^{+8}_{-7}$ & 42(4) & 39(2) \\
\hline
\multicolumn{9}{c}{$\bullet$ $\beta=7.0$, scaled by $a_\rho$ $\bullet$} \\
$k \leq 1$ & 1760(100) & $102^{+6}_{-11}$ & 91(6) & 84(4)
           & 4500(400) & $79^{+7}_{-6}$ & 36(3) & 34(2) \\
$k \leq 2$ & 1750(90) & $103^{+6}_{-9}$ & 104(6) & 96(6)
           & 4500(400) & $79^{+7}_{-6}$ & 39(4) & 37(2) \\
$k \leq 3$ & 1880(90) & $103^{+7}_{-8}$ & 101(6) & 91(4)
           & 4500(400) & $79^{+7}_{-6}$ & 39(3) & 36(2) \\
\end{tabular}
\end{table*}

At $\beta=6.8$, simulations up to $O(1/M)$, $O(1/M^2)$ and $O(1/M^3)$ each
produce masses for both the $D_s$ and $B_s$ which
are consistent with the experimental values, indicating that the bare charm
and bottom masses from quarkonium physics are also relevant to
heavy-light mesons. 
The light quark dependence at $\beta=6.8$, as probed by $D_s-D$ and $B_s-B$, 
is also in reasonable agreement with experiment.  
The spin splittings are significantly
smaller than experiment, which is a general feature of previous lattice 
results as well\cite{review,Japan,previous,ADCSS,melbourne}, 
and is often attributed to quenching.

The situation at $\beta=7.0$ is complicated by the fact that $a_\rho \neq
a_{\rm hvy}$.  In Table \ref{tab:phys}, the lattice data are shown 
for both of these normalizations with the bare heavy quark masses fixed to
the values obtained from quarkonium.  The use of $a_{\rm hvy}$ produces 
$D_s$ and $B_s$ masses which agree nicely with experiment (as was found for
$\beta=6.8$), whereas the data normalized to $a_\rho$ clearly cannot produce
accurate $D_s$ and $B_s$ masses when the bare masses are fixed by quarkonium.

Conversely, mass differences normalized to $a_{\rm hvy}$ tend to be larger 
than the results at $\beta=6.8$, whereas the $a_\rho$-normalized mass
differences are found to scale remarkably well with respect to the 
$\beta=6.8$ results.
This preference of the data for $a_\rho$ is in accordance with the familiar 
notion that the dynamics of heavy-light mesons is governed by the light 
degrees of freedom, rather than by explicit heavy quark dynamics.

Perhaps the most satisfactory determination of mass differences at $\beta=7.0$
would be obtained by normalizing to $a_\rho$ and re-tuning the bare mass
to the heavy-light spectrum itself, with no reference to quarkonium.
However, the possibly-problematic convergence of the $1/M$ expansion for 
charmed mesons precludes a more detailed effort in this direction at present.
There is no re-tuning required at $\beta=6.8$, so at least in this case
an unambiguous quantitative comparison to experiment can be made from
the data in Table \ref{tab:phys}, although concerns about the potentially-large 
$O(1/M^3)$ contributions (such as the $c_7$ term) must certainly be addressed.

In the work of
Ishikawa {\it et al.}\cite{Japan}, a range of heavy quark masses were
studied and it was found that the $O(1/M^2)$ terms
increase the spin splitting relative to the $O(1/M)$ value, in agreement
with what is reported here in Table \ref{tab:phys}.
The classically-complete set of $O(1/M^3)$ terms have now been included, and
their total contribution is smaller in magnitude than $O(1/M^2)$.
However, it has also been demonstrated that the small $O(1/M^3)$ contribution
to the spin splitting results from a cancellation involving an individual
term (the $c_7$ term) which is dangerously-larger in magnitude than $O(1/M^2)$.

\section{CONCLUSIONS}

The masses of $^1S_0$ and $^3S_1$ heavy-light mesons have been obtained
from quenched lattice NRQCD at two lattice spacings, near 0.22fm and 0.26fm,
using tadpole-improved, classically-improved light quark and gauge field
actions.
Results were obtained separately at $O(1/M)$, $O(1/M^2)$ and $O(1/M^3)$.
The effects of individual terms at $O(1/M^3)$ were also shown.

The simulations up to $O(1/M^2)$ support the existing knowledge of heavy-light
S-waves for lattice QCD.  Masses are in qualitative agreement with experimental
data, except that the spin splitting is noticeably smaller than experiment.
This may be due, at least in part, to quenching.

The contributions of the $O(1/M^3)$ terms have not been studied in detail
previously.  A novel feature at this order is the existence of a large
vacuum expectation value (in the $c_{10}$ term) that shifts the heavy quark 
mass.  Special care
is needed when this vacuum value is present, particularly in the charm region.
An important example is Fig. \ref{fig:hypord3}, where the vacuum value seems
to make a sizeable contribution to the S-wave spin splitting, but the effect
was shown to be an artifact of the familiar discretization of $\exp(H\tau)$ in
the heavy quark propagation.

It should be noted that the $c_7$ term also produces an $O(1/M^3)$ contribution 
to the spin splitting which is larger in magnitude than the total 
$O(1/M^2)$ piece.  In fact, Fig. \ref{fig:hypord3} indicates that this effect 
is very similar in size to the spurious contribution from the vacuum value
in the $c_{10}$ term, 
thus raising the question of whether the $c_7$ contribution might also 
contain some artifact related to the discretization of $\exp(H\tau)$.
This systematic uncertainty has not been discussed in the literature to date, 
but might offer some important insight for the charmed spectrum calculation
and also for charmonium.

Other possibilities for reducing the magnitude of the $O(1/M^3)$ corrections 
also deserve further study.  In the present work, the coefficients of the 
NRQCD Hamiltonian
have been approximated by their classical values, along with tadpole 
improvement.  It would be interesting to see the effects of
retaining one-loop perturbative or nonperturbative renormalization for these 
coefficients.  One might
also consider working at a smaller lattice spacing, although this will
move the charm quark mass even further away from the heavy quark limit.  
(Recall that $aM_c = 1.1$ at
$\beta = 7.0$.  According to Ref. \cite{Trottier}, $aM_c = 0.81$ at
$\beta = 7.2$.)  Some benefit might come from using a different definition
of the tadpole factor, such as the Landau link definition, which
increases $aM_c$ at a fixed lattice spacing.\cite{Trottier,Shake}

The existence of an indivdual $O(1/M^3)$ term whose contribution to the 
spin splitting is larger than the cumulative
$O(1/M^2)$ effects, as was found in the present work, indicates that
the application of lattice NRQCD to charmed mesons requires care.

\acknowledgments

The authors are grateful to Howard Trottier for useful discussions and for
access to his NRQCD codes, and to G. Peter Lepage for an important 
communication regarding the large vacuum expectation value at $O(1/M^3)$.
This work was supported in part by the Natural Sciences and Engineering
Research Council of Canada.  R.L. also acknowledges support from 
the U.S. Department of Energy, contract DE-AC05-84ER40150.

\vfil\eject

\end{document}